\documentstyle[prl,aps,epsf,multicol]{revtex}   %Comment out for Latex209

\begin{document}
\draft

\title{Defect-Unbinding Transition in Layered Superconductors}

\author{Matthew J.W.\ Dodgson, Vadim B.\ Geshkenbein, and Gianni Blatter}

\address{Theoretische Physik, ETH-H\"onggerberg, CH-8093
  Z\"urich, Switzerland}

%\twocolumn[                   %Comment out for new Latex
%\date{\today}
\date{February 17, 1999}
\maketitle
%\widetext                     %Comment out for new Latex
%\vspace*{-1.0truecm}          %Comment out for new Latex
\begin{abstract}
%\begin{center}                %Comment out for new Latex
%\parbox{14cm}{                %Comment out for new Latex
  
We establish a new interstitial--vacancy unbinding transition 
of the Berezinskii-Kosterlitz-Thouless type, transforming the three 
dimensional pancake vortex lattice of a decoupled layered 
superconductor into a defected solid. This transition is 
the natural finite-field extension of the vortex--anti-vortex 
unbinding transition establishing the zero-field superfluid 
stiffness. At finite Josephson coupling, the defect unbinding 
transition turns into a topological decoupling transition.
%}                            %Comment out for new Latex

%\end{center}                 %Comment out for new Latex

\end{abstract}
%]                            %Comment out for new Latex


\begin{multicols}{2}          %Comment out for Latex209
\narrowtext

The soft vortex matter in high-temperature layered 
superconductors exhibits a fascinating rich phase diagram 
with a variety of phase transitions modifying the various 
intra- and interplanar correlations.  These include
a melting transition from a line solid to a line liquid 
\cite{Nelson,Houghton,review}, a sublimation transition taking the solid 
into a pancake vortex gas \cite{BGLN}, and a decoupling transition 
\cite{BGLN,GlazmanKoshelev,Daemon,HorovitzGoldin}
destroying the superconducting coherence between 
the planes. In this letter we concentrate on the 
decoupled limit where the three-dimensional (3D) nature
of the vortex system is solely due to the long range 
electromagnetic interaction between the pancake vortices.
We establish that in the crystal phase there is a
Berezinskii-Kosterlitz-Thouless (BKT) phase transition which separates the 
pancake-vortex lattice from a defected 3D solid with a 
finite density of free interstitials and vacancies, see Fig.~1. 
With finite Josephson interactions this
defect-unbinding transition also triggers the decoupling of the 
layers, preempting other proposed mechanisms of 
decoupling \cite{Daemon,HorovitzGoldin}.
\begin{figure} [bt]
\vspace{-0.1cm} 
\centerline{\epsfxsize=7.5cm\epsfbox{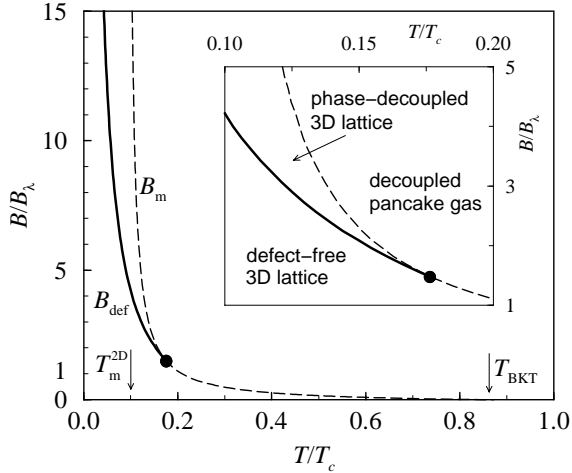}}
\narrowtext
\caption{Phase diagram for a weakly-coupled layered
superconductor. The solid line marks the defect-unbinding 
transition $B_{\rm def}(T)$ which transforms into the decoupling 
transition $B_{\rm dec}(T)$ in the case of a finite weak Josephson 
coupling with $\varepsilon < d/\lambda$. The dashed line shows the melting
line $B_{\rm m}(T)$ calculated within a self-consistent analysis.
(We have used parameters $\lambda(0) \approx$
2000~\AA, $d\approx$ 15~\AA, and $T_c \approx$ 100~K.)}    
\end{figure}

The vortex lattice in layered superconductors is ``soft'' and
melts at low temperatures; for fields $B>B_\lambda =
\Phi_0/\lambda^2$ ($\lambda=$ penetration depth) the transition 
is governed by the shear interaction between pancake vortices
in the same layer and thus is close to the two-dimensional 
dislocation-mediated BKT melting in the individual layers 
\cite{Fisher}. The 3D nature of the transition becomes explicit 
only at low fields $B<B_\lambda$, where the electromagnetic 
(tilt) interaction between pancake vortices in different layers 
dominates over the intralayer (shear) interaction due to the planar 
currents \cite{BGLN}. Here, we study another thermodynamic property
of the pancake-vortex lattice -- its susceptibility to the 
formation of free defects in the form of pancake-vortex interstitials 
and vacancies. It has recently been shown \cite{SlutzkyBrandt}
for the pancake-vortex lattice in layered decoupled superconductors 
that the self-energy of (and the logarithmic interaction 
between) vacancies and interstitials is strongly screened 
due to the relaxation of the vortex lattice surrounding the defects. 
We find that the corresponding interstitial--vacancy pair 
creation can be understood in terms of a simple vortex--anti-vortex 
pair creation in a superconductor with a vortex-induced suppression 
of the superfluid density. The usual zero-field BKT
transition establishing the superfluid stiffness of the individual 
superconducting layers \cite{KosterlitzThouless} therefore has its counterpart 
at finite fields in the form of interstitial--vacancy unbinding
in the vortex solid. 

We start from a rigid vortex lattice in a decoupled layered 
superconductor and move one pancake by the distance $R$ within 
the same layer to produce an interstitial--vacancy defect pair. 
The excitation energy associated with this manipulation 
contains a self-, an interaction-, and a core-energy part and reads
\begin{equation}
E_{i,v}(R,L) = 2\varepsilon_0 d \left[\ln\frac{L}{a_0} - \ln\frac{L}{R}\right]
+(\eta_i +\eta_v) \varepsilon_0 d.
\label{Energy_iv}
\end{equation}
Here, $\varepsilon_0 = (\Phi_0/4 \pi\lambda)^2$ is the basic line
energy associated with the vortex, $d$ is the layer separation, 
$\lambda$ denotes the planar penetration depth, $a_0 = \sqrt{\Phi_0/B}$ 
is the lattice constant, and $L$ the macroscopic extent of the layer. 
The numericals $\eta_{i,v}$ quantify the core energies of the 
interstitial- and the vacancy-defect and are of order 0.1 
\cite{Frey,OliveBrandt}. 

Next, we let the lattice relax around the defect. We concentrate 
on a single interstitial and determine the screened self-energy -- 
the interaction between the defects is screened in the same 
manner. We first ignore the electromagnetic coupling to the 
pancake vortices in the other layers and take them into account 
in a second step later. The interaction energy of the 
pancake-vortex configuration reads
\[
E_{\rm int} = \frac{1}{2}\sum_{i,j} 
V_{\rm pc}({\bf R}_i - {\bf R}_j) 
= \frac{1}{2} \int_{\rm\scriptscriptstyle BZ} 
\frac{d^2 K}{(2\pi)^2} \, n_{\bf\scriptscriptstyle K} 
V_{\rm pc}({\bf K}) \, n_{-{\bf\scriptscriptstyle K}}
\]
($V_{\rm pc} (R) = - 2\varepsilon_0 d \ln(R/a_0)$ is the
pancake interaction energy and $n_{\bf\scriptscriptstyle K} = 
\sum_i e^{i{\bf K}.{\bf R}_i}$ the Fourier transform of the 
pancake-vortex density), from which one can derive the dispersive 
compression modulus 
\begin{eqnarray}
c_{11}(K) = n^2 V_{\rm pc} (K) = \frac{4\pi\varepsilon_0 d}{a_0^4 K^2},
\label{c11}
\end{eqnarray}
with $n = B/\Phi_0$ the vortex density. The elastic energy of 
the lattice with one defect can be written in the form
\begin{equation}
E_{\rm el} = \!\int_{\rm\scriptscriptstyle BZ} \!\frac{d^2K}{(2\pi)^2}
\Big[\frac{c_{11}}{2} |{\bf K}\cdot{\bf u}_{\bf\scriptscriptstyle K}|^2 
+ i n ({\bf K}\cdot{\bf u}_{\bf\scriptscriptstyle K})\, 
V_{\rm pc}(K) \Big],
\label{E_elast}
\end{equation}
where ${\bf u}_i$ denotes the displacement field of the pancake vortex
lattice, ${\bf u}_{\bf\scriptscriptstyle K}$ its Fourier transform. While the first
term in (\ref{E_elast}) describes the elastic energy of the relaxed lattice, the
second term originates from the force the interstitial exerts on the
pancake vortices, $ E_{\rm source} =\sum_i V({\bf R}_i -{\bf u}_i)$
(we place the interstitial at the origin). Minimizing (\ref{E_elast})
with respect to ${\bf u}$ we obtain the distortion 
\begin{equation}
{\bf u}_{\bf\scriptscriptstyle K} = - \frac{i n {\bf K} V_{\rm pc} (K)}
{c_{11}(K) K^2} = -\frac{i}{n}\frac{{\bf K}}{K^2},
\label{displacement}
\end{equation}
where we have made use of (\ref{c11}).  The displacement 
${\bf u}({\bf R})={\bf R}/2\pi n R^2$ removes exactly one
pancake vortex from the neighborhood of the interstitial and thus
produces perfect screening \cite{SlutzkyBrandt}. As a consequence, the 
log-divergent self-energy in (\ref{Energy_iv}) is compensated 
by the gain in elastic energy under the relaxation of the lattice: 
Inserting the solution (\ref{displacement}) back into (\ref{E_elast}), 
we obtain
\begin{equation}
E_{\rm el} = - \frac{1}{2} \int_{\rm\scriptscriptstyle BZ} 
\frac{d^2K}{(2 \pi)^2} V_{\rm pc}(K) = - \varepsilon_0 d \ln \frac{L}{a_0}.
\label{E_elast_f}
\end{equation}
This perfect screening is spoiled by the interaction of the pancake
vortices in the layer ($n=0$) with those in the other layers 
($n \ne 0$). Let us view the lattice as being made from columns
of pancake vortices. The electromagnetic interaction of the pancake
vortices in the $n=0$-layer with the rest of the lattice then can be 
cast into the form of a substrate potential
\begin{equation}
V_{\rm sub} = \frac{1}{2} \sum_i \alpha_s u_{n=0,i}^2,
\label{V_sub}
\end{equation}
where the curvature $\alpha_s = (\varepsilon_0d/\lambda^2) \ln (a_0/d)$
follows from the electromagnetic energy associated with displacing 
a single vortex in a column \cite{Feigelman,cutoff}. While the individual
contribution from each pancake vortex in the column involves the
prefactor $d/\lambda$ and hence is small, the long-range nature of the
electromagnetic interaction leads to a large number ($\sim \lambda/d$)
of terms from distant layers. Adding the substrate energy 
$V_{\rm sub}$ to the elastic energy (\ref{E_elast}) and minimizing,
the displacement field (\ref{displacement}) is reduced by the 
factor $1+g$ with \cite{SlutzkyBrandt}
\begin{equation}
g = \frac{n \alpha_s}{c_{11}(K) K^2} = \frac{a_0^2}{4\pi\lambda^2}
\ln \frac{a_0}{d}\approx \frac{H_{c1}}{B}.
\label{g}
\end{equation}
($H_{c_1}$ is the lower critical field.)
The screening charge is reduced to the fraction $1/(1+g)$ of a pancake
vortex and correspondingly the divergent self-energy in
(\ref{Energy_iv}) is only partially cancelled by the lattice
relaxation; inserting ${\bf u}_{\bf\scriptscriptstyle K}$ back into the
expression for the elastic energy, we obtain $E_{\rm el}=-\varepsilon_0 
d \ln (L/a_0)/(1+g)$. The relaxation of the pancake-vortex lattice then 
reduces the self-energy in (\ref{Energy_iv}) by the factor 
$g/(1+g)$. At large fields, $B > H_{c_1}$, where the vortex-vortex 
interaction is dominated by the intraplanar forces we have $g<1$ and 
find a large downward renormalization of the energy scale of the defects.  
When $g>1$ at small fields $B < H_{c_1}$, the interaction between the pancake 
vortices is dominated by the interlayer forces and the effect of screening is 
small.

The above analysis extends trivially to the case of a pancake-vortex
vacancy as well as to the interaction between the defects. 
The energy for an interstitial--vacancy pancake-vortex pair of 
extent $R$ placed into one plane of a decoupled layered 
superconductor then involves the screened self- and interaction-energies 
\begin{equation}
E_{i,v}^{\rm sc} (R,L) =  \frac{2 g \varepsilon_0 d}{1+g} 
\ln\frac{R}{a_0} + 
(\eta_i'+\eta_v') \varepsilon_0 d.
\label{Energy_iv_sc}
\end{equation}
Before proceeding, we comment on the approximation we have made 
in determining the substrate potential (\ref{V_sub}), where we
have ignored the relaxation of the pancake-vortex positions in
neighboring layers: a simple estimate shows that the relaxation 
$u_{n,i}$ of the pancakes in the \hbox{$n$-th} layer is smaller by the factor 
$d/\lambda$, $u_{n,i} \sim (d/\lambda) u_{0,i}$, and hence the correction 
to the substrate potential (\ref{V_sub}) is small, $\delta V_{{\rm sub},i} 
\sim (\lambda/d)\alpha_s u_{n,i}^2 \sim (d/\lambda)\alpha_s u_{0,i}^2$.

We now discuss the implications of the result (\ref{Energy_iv_sc}) 
for the pancake-vortex phase diagram. An immediate consequence of 
the logarithmic interaction between the defects is a
Berezinskii-Kosterlitz-Thouless transition \cite{KosterlitzThouless}.
The field dependent screening of the 
interaction pushes this defect-unbinding transition to low 
temperatures,
\begin{equation}
T_{\rm def} (B) = \frac{g}{1+g} \frac{\varepsilon_0 d}{2},
\label{T_def}
\end{equation}
and solving for $B$ we obtain the defect-unbinding line
\begin{equation}
B_{\rm def} (T) = \frac{\Phi_0}{4 \pi \lambda^2} 
\ln\Bigl(\frac{a_0}{d}\Bigr) \left( \frac{\varepsilon_0 d}{2T} -1 \right).
\label{B_def}
\end{equation}
In the presence of a finite Josephson coupling, the appearance of free
defects in the solid triggers the decoupling of the superconducting
layers and the defect-unbinding line $B_{\rm def} (T)$ transforms to a 
decoupling line $B_{\rm dec} (T)$. For fields above
$B_{\rm dec} (T)$ the system develops a finite $c$-axis resistivity
with $\rho_c$ proportional to the number of free mobile defects
$n_d\propto a_0^{-2}
\exp(-2b/\sqrt{1-T/T_{\rm def}})$,
see Ref.\ \onlinecite{Koshelev} ($b$ is a non-universal constant). 
This topological decoupling 
transition corresponds to the quartet-unbinding transition 
first proposed by Feigel'man {\it et al.} \cite{Feigelman} (or the supersolid
transition \cite{Frey}; a defect proliferation transition is also described in 
\cite{CarruzzoYu})--- 
here, we establish a rigorous basis for the 
transition via the BKT scenario, accounting for the screening response 
of the pancake-vortex lattice.
A finite Josephson coupling does not change
result (\ref{T_def}) if
the extra Josephson 
energy $\xi_{\rm def}^2 E_{\rm\scriptscriptstyle J}$
within the coherence area \hbox{$\xi_{\rm def}^2 \sim n_d^{-1}$}
remains small, hence $a_0^2 
\varepsilon^2 \varepsilon_0/d < T_{\rm def} \sim g \varepsilon_0 d$, 
producing the condition $\varepsilon < d/\lambda$.

In the past, a number of scenarios have been proposed for the
decoupling transition in layered superconductors. We have to
distinguish between weak (\hbox{$\lambda < \Lambda$}) and intermediate
(\hbox{$\Lambda < \lambda$}) coupling, where we have defined the Josephson
screening length \hbox{$\Lambda = d/\varepsilon$}. For weak coupling we
define the crossover field 
\hbox{$B_\times = B_\lambda \Lambda^2/\lambda^2>B_\lambda$}. 
Below $B_\times$, phase fluctuations are determined by
the interlayer (tilt) interaction and the
decoupling line has been predicted to take the form $B_{\rm dec} (T) = 
B_\lambda (\varepsilon_0 d/\pi T) \ln(a_0/d)$
\cite{Daemon,HorovitzGoldin}. Above $B_\times$, the intralayer (shear)
interaction is important
and the decoupling line takes the form $B_{\rm dec} (T) \sim B_\Lambda
(\varepsilon_0 d/T)^2$ \cite{BGLN}. For an intermediate
coupling the crossover field takes the form $B_\times = B_\Lambda =
\Phi_0/\Lambda^2$ and the low and high field decoupling lines read
$B_{\rm dec} (T) = B_\Lambda (\varepsilon_0 d/e \pi T) < B_\times$
\cite{Daemon} and $B_{\rm dec} (T) \sim B_\Lambda (\varepsilon_0
d/T)^2 > B_\times$ \cite{GlazmanKoshelev}. Parametrically, all these
results are easily reproduced via a Lindemann criterion for the
interlayer phase correlator \cite{BGLN,GlazmanKoshelev}. 
Our result for topological decoupling (\ref{B_def}) is
valid for weak coupling below $B_\times$ and so should
be compared to the results of Daemen {\it et al.}
\cite{Daemon} and of Horovitz and Goldin \cite{HorovitzGoldin}. 
Of course, whichever decoupling transition occurs first can be the only such
transition, and we find that our topological decoupling line
$B_{\rm dec} (T)$ is a factor 8 below the line advocated in
Refs.\ \cite{Daemon,HorovitzGoldin}. 
This numerical relation
between the two results is not fortuitous, as we will now discuss.

In fact, the result (\ref{Energy_iv_sc}) and the associated
transition (\ref{T_def}) is easily understood as the finite-field
extrapolation of the zero-field vortex--anti-vortex pair creation and
their unbinding transition. This understanding brings forward an
interesting parallel between the competing zero-field intralayer
vortex unbinding- \cite{Feigelman} and the interlayer vortex loop 
transition suggested by Friedel \cite{Friedel} on the one hand, and the present
intralayer defect unbinding- and the interlayer decoupling transition 
advocated by Daemen {\it et al.}~and by Horovitz and Goldin 
on the other hand. In order to understand that we deal with the 
identical physics in both cases we start from the London free energy 
of a layered superconductor, ${\cal F}=\sum_n F_n+\int d^3 r 
\, B^2/8\pi$ with
\begin{equation}
F_n = \! \int \!\! d^2R \, \frac{\varepsilon_0 d}{2\pi} \Bigl[ (\nabla 
\varphi_n + {\bf a})^2 \! + \!\frac{2\varepsilon^2}{d^2}
(1\!-\!\cos \phi_{n,n+1})\Bigr].
\label{F_n}
\end{equation}
Here, $\nabla = ({\partial_x,\partial_y})$, ${\bf a} = 2 \pi {\bf
A}/\Phi_0$ is the planar component of the vector potential,
$\phi_{n,n+1} = \varphi_{n+1}-\varphi_n + \int dz a_z$, and
$\varepsilon^2 = m/M < 1$ is the mass anisotropy ratio. For vanishing
interlayer coupling, the individual layers undergo a zero-field
BKT vortex unbinding transition at $T_{\rm\scriptscriptstyle BKT} 
= \varepsilon_0 d/2$ \cite{Feigelman}. On the other hand,
it has been suggested by Friedel \cite{Friedel}, that the coupled system might
undergo a decoupling transition triggered by low-energy interplanar
vortex loops. However, as shown by Korshunov \cite{Korshunov}, this
transition would take place only at higher temperatures
$T_{\rm\scriptscriptstyle L} = 4 \varepsilon_0 d$ and is therefore
preempted by the vortex unbinding transition at
$T_{\rm\scriptscriptstyle BKT}$.

In finite fields the presence of vortices has to be accounted for.  A
simple derivation of the appropriate free energy starts from the
continuous elastic description of the pancake-vortex system, combining
compression, shear, and tilt energies. In the incompressible limit we
can drop the compression term and write the phase variable in terms of
the displacement field ${\bf u}$, $\varphi_{\bf\scriptscriptstyle K} 
= (2\pi i/a_0^2) ({\bf u}_{\bf\scriptscriptstyle K}\times{\bf K})/K^2$. 
The electromagnetic tilt energy takes the form 
$g\, \varepsilon_0 d \, (\nabla \varphi)^2/2\pi$ with $g$
given by (\ref{g}). Comparing with (\ref{F_n}), the
factor $g$ accounts for the suppression of the superfluid density due
to the presence of the vortices. In order to obtain the correct limit
for $B \rightarrow 0$ we have to include the bare superfluid
density. This is easily achieved via a calculation of the effective
penetration depth $\lambda_{\rm eff}$ in the presence of vortices,
$\lambda_{\rm eff}^2 = \lambda^2 + B^2 / 4\pi
\alpha_{\rm\scriptscriptstyle L}$, where $\alpha_{\rm\scriptscriptstyle
L}$ denotes the Labusch parameter describing vortex ``pinning,'' $V_{\rm
pin,i} = \alpha_{\rm\scriptscriptstyle L} u_i^2/2$. Here, ``pinning'' is
due to the substrate potential (\ref{V_sub}) and hence $\alpha_{\rm
\scriptscriptstyle L} = \alpha_s/a_0^2 d = 4\pi \varepsilon_0 g/a_0^4$. 
Collecting results, we obtain the suppression factor for the superfluid 
density $\rho_{\rm eff} / \rho = \lambda^2/\lambda_{\rm eff}^2 = 
g/(1+g)$. In the end, the free energy replacing (\ref{F_n}) in 
the presence of vortices takes the form
\begin{eqnarray}
F_n &=& \! \int \! d^2R \,\frac{\varepsilon_0 d}{2\pi} \,
\Bigl[ \frac{g}{1+g} (\nabla \varphi_n + {\bf a})^2 + \frac{a_0^4}{16\pi}
(\Delta\varphi_n)^2 \nonumber \\
&+&  \frac{2\varepsilon^2}{d^2}(1-\cos \phi_{n,n+1})\Bigr].
\label{F_n_B}
\end{eqnarray}
For fields $B < B_\times = B_\lambda \Lambda^2/\lambda^2$ we can drop the
shear term $\propto (\Delta \varphi)^2$ in comparison with the
electromagnetic and Josephson tilt energies. The free energy
(\ref{F_n_B}) then is identical in form to the one studied by
Korshunov \cite{Korshunov}; correspondingly, we can discuss two
transitions: Turning the phase by $\pm 2\pi$ we add/remove a vortex
from the system and thus recover the energy (\ref{Energy_iv_sc}) for
the interstitial--vacancy pair creation and the corresponding unbinding
transition as given by (\ref{T_def}). On the other hand, Friedel's
loop transition takes place at a temperature $T_{\scriptscriptstyle L}
(B) = 4 [g/(1+g)] \varepsilon_0 d$, which is just the result obtained by
Daemen {\it et al.} \cite{Daemon} and by Horovitz 
and Goldin \cite{HorovitzGoldin}. 

The pair of transitions discussed above is not unique to the vortex
system discussed here --- it appears in the context of the $XY$-model
(with coupling $J$) subject to a symmetry breaking field $\propto 
\cos[m \vartheta_{\bf\scriptscriptstyle R}]$ (here, $\vartheta_{\bf
\scriptscriptstyle R}$ denotes the angle of the spin at the position 
${\bf R}$, $m$ is an integer), where the BKT transition at 
$T_{\rm\scriptscriptstyle BKT} = \pi J/2$ competes with the unlocking
or `roughening' transition at $T_{\rm\scriptscriptstyle R} = 8\pi J/m^2$
\cite{Jose}. A similar scenario shows up in the context of the 
adsorption of a 2D crystal (elasticity $C$, lattice constant $a$) 
on a commensurate substrate (lattice constant $a/m$), where the 
dislocation mediated melting transition at $T_{\rm m} = C a^2/4 \pi$ competes 
with the depinning or `roughening' transition at $T_{\rm\scriptscriptstyle R} = 
4 C a^2/\pi m^2$ \cite{PokrovskyTalapov} (in both cases the ratio
between the two transition temperatures involves the factor $16/m^2$,
twice larger than in the present layered situation). In that sense, 
Friedel's loop transition corresponds to the ($m=1$) roughening 
transition which is preempted by a topological transition at lower 
temperatures.

%As discussed before, above the crossover field $B_\times = B_\lambda 
%\Lambda^2/\lambda^2$ the result (\ref{B_def}) has to be replaced by 
%the shear dominated expression $B_{\rm dec} \approx B_\Lambda$ 
%$[\ln(\varepsilon\lambda^2/d^2)/8\pi G^{\rm\scriptscriptstyle 2D}]^2
%(T_c/T)^2$. 
%On the other hand, 

Thermal fluctuations will soften the substrate potential and modify the result
(\ref{B_def}): At high 
temperatures and low fields the renormalized substrate potential 
(\ref{V_sub}) can be obtained from the thermal average $\alpha_s(T) =
\sum_n \langle \partial^2 V_n({\bf u}_n-{\bf u}_0)/(\partial u_{0,x})^2
\rangle_{\rm th}$ (see Ref.\ \cite{Dodgson} for details). 
Going over to Fourier space, the ${\bf K}$ component of 
the substrate potential is smoothed by the Debye-Waller factor 
$\exp[-K^2 \langle u^2\rangle_{\rm th}/2]$. The mean squared 
displacement is in turn determined through the substrate potential, 
$\langle u^2\rangle_{\rm th} = 2T/\alpha_s$. Solving self-consistently,
we find $\langle u^2\rangle_{\rm th} = 2\lambda^2/[(\varepsilon_0 d/2T)-1)$ 
and $\alpha_s(T) = [\varepsilon_0 d/\lambda^2(0)] 
(1-T/T_{\rm\scriptscriptstyle BKT})$. Here, we have used a mean-field 
temperature dependence $\lambda^2(T) = \lambda^2(0)/(1-T^2/T_c^2)$, which
gives a zero-field BKT transition at $T_{\rm\scriptscriptstyle
BKT} =\varepsilon_0(T_{\rm\scriptscriptstyle BKT})$ $d/2 \Longrightarrow
T_{\rm \scriptscriptstyle BKT} = \varepsilon_0(0)d/
(1+\varepsilon_0(0)d/T_c)$. The decoupling line then has to be 
determined from (\ref{T_def}) using the renormalized suppression factor
$g(T) = a_0^2(1-T/T_{\rm\scriptscriptstyle BKT})/4\pi\lambda(0)^2$ for the
superfluid density and we obtain the final result near 
$T_{\rm\scriptscriptstyle BKT}$
\begin{equation}
B^r_{\rm def} (T) = \frac{\Phi_0}{4\pi \lambda^2 (0)}
\frac{\varepsilon_0(0) d}{T} 
\left(1 - \frac{T}{T_{\rm \scriptscriptstyle BKT}}\right)^2.
\label{B_def_r}
\end{equation}
The position of the line $B^r_{\rm def} (T)$ should be compared
with the melting (or sublimation) line $B_{\rm m}(T)$. Using the above 
self-consistent 
analysis we have determined the line in the $B$--$T$ phase diagram where 
the substrate potential collapses and the 3D vortex solid becomes 
unstable \cite{Dodgson}. The numerical result is shown as the dashed line in
Fig.~1, interpolating between $T_{\rm\scriptscriptstyle BKT}$ at zero field and
the 2D melting
temperature $T_{\rm m}^{\rm 2D}\approx  \varepsilon_0 d / 70$ at high fields.
The asymptotic limit at low fields takes the form
$B_{\rm m}(T) = (B_\lambda/16\pi)
(T_{\rm \scriptscriptstyle BKT}/T)^2 (1-T/T_{\rm \scriptscriptstyle BKT})^2$.
Comparing this with (\ref{B_def_r}) we find that the melting
line undercuts the defect-unbinding line at high temperatures, leaving
only one (sublimation) transition at $B_{\rm m}(T)$. At low temperatures, 
$B_{\rm def} (T)$ is below the melting line,
allowing for two transitions separating 
a defect-free solid, a phase-decoupled solid and a pancake gas. 
The defect-unbinding line 
$B^r_{\rm def} (T)$ calculated with the self-consistently softened substrate is
shown as the full line in Fig.~1.

Though screening strongly reduces the interaction energy 
of the defects for fields $B > B_\lambda$, their core energy remains 
high. For an estimate we consult the numerical work of Frey 
{\it et al.} \cite{Frey} and of Olive and Brandt \cite{OliveBrandt},
who find values of order $0.15 - 0.2 \varepsilon_0$ for the 
vacancy and interstitial line defects in a 3D vortex lattice, the 
interstitial typically being 30 \% cheaper in energy than the vacancy.
Therefore at high fields the defect density at the unbinding transition
is small, $n_d \propto a_0^{-2} \exp [-\lambda^2/a_0^2]$, (with more
interstitials than vacancies) and the 
transition is weak.

In conclusion, we have established a rigorous framework for
the topological decoupling transition in layered type II 
superconductors which is based on a defect-unbinding transition 
of the BKT type.

We thank the Fonds National Suisse for financial support.

\end{multicols}          %Comment out for Latex209


\vspace{-0.5cm} 

\begin{thebibliography}{99}

\vspace{-1.5truecm}

\bibitem{Nelson} D.R.\ Nelson, 
Phys.\ Rev.\ Lett.\ {\bf 60}, 1973 (1988).

\bibitem{Houghton} A.\ Houghton, R.A.\ Pelcovits, and A. Sudb\o,
Phys.\ Rev.\ B {\bf 40}, 6763 (1989).


\bibitem{review} G. Blatter {\it et al.}, 
Rev.\ Mod.\ Phys.\ {\bf 66}, 1125 (1994).

\bibitem{BGLN} G.\ Blatter {\it et al.}, 
%V.B.\ Geshkenbein, A.I.\ Larkin, and H.\ Nordborg, 
Phys.\ Rev.\ B {\bf 54}, 72 (1996).

\bibitem{GlazmanKoshelev} L.\ Glazman and A.E.\ Koshelev,
Phys.\ Rev.\ B {\bf 43}, 2835 (1991).

\bibitem{Daemon} L.\ Daemen {\it et al.}, 
%L.\ Bulaevskii, M.\ Maley, and J.\ Coulter,
Phys.\ Rev.\ Lett.\ {\bf 70}, 1167 (1993).

\bibitem{HorovitzGoldin} B.\ Horovitz and T.R.\ Goldin,
Phys.\ Rev.\ Lett.\ {\bf 80}, 1734 (1998).

%\bibitem{HubermanDoniach} B.\ Huberman and S.\ Doniach,
%Phys.\ Rev.\ Lett.\ {\bf 43}, 950 (1979).

\bibitem{Fisher} D.\ Fisher,
Phys.\ Rev.\ B {\bf 22}, 1190 (1980).

\bibitem{SlutzkyBrandt} M.\ Slutzky, R.\ Mints, and E.H.\ Brandt,
Phys.\ Rev.\ B {\bf 56}, 453 (1997); E.H.\ Brandt, 
Phys.\ Rev.\ B {\bf 56}, 9071 (1997).

\bibitem{KosterlitzThouless} V.L.\ Berezinskii, Sov. Phys. JETP 
{\bf 32}, 493 (1971); Sov. Phys. JETP {\bf 34}, 610 (1972);
J.M.\ Kosterlitz and D.J.\ Thouless, 
J.\ Phys.\ C {\bf 6}, 1181 (1973).

\bibitem{Frey} E.\ Frey, D.R.\ Nelson, and D.S.\ Fisher,
Phys.\ Rev.\ B {\bf 49}, 9723 (1994).

\bibitem{OliveBrandt} E.\ Olive and E.H.\ Brandt,
Phys.\ Rev.\ B {\bf 57}, 13861 (1998).

%\bibitem{strictly_valid} This expression for the compression modulus
%is strictly valid only for a long range interaction $V_{\rm pc} (R)$
%between the pancake vortices. 

\bibitem{Feigelman} M.V.\ Feigel'man, V.B.\ Geshkenbein, and A.I.\ Larkin,
Physica C {\bf 167}, 177 (1990).

\bibitem{cutoff} At low fields, where $a_0 > \lambda$ the lattice
constant $a_0$ under the logarithm has to be replaced by the
penetration depth $\lambda$.

\bibitem{Koshelev} A.~E.\ Koshelev, 
Phys.\ Rev.\ Lett.\ {\bf 76}, 1340 (1996). 

\bibitem{CarruzzoYu} H.M.\ Carruzzo and C.C.\ Yu, Phil.\ Mag.\ B {\bf 77},
  1001 (1998).

\bibitem{Friedel} J.\ Friedel,
J.\ Phys.\ France {\bf 49}, 1561 (1988).

\bibitem{Korshunov} S.E.\ Korshunov,
Europhys.\ Lett. {\bf 11}, 757 (1990).

\bibitem{Jose} J.V.\ Jos\'e {\it et al.}, 
Phys.\ Rev.\ B {\bf 16}, 1217 (1977).

\bibitem{PokrovskyTalapov} V.L.\ Pokrovsky and A.L.\ Talapov,
Phys.\ Rev.\ Lett.\ {\bf 42}, 65 (1979).

\bibitem{Dodgson} M.J.W.\ Dodgson {\it et al.}, in preparation.

\end{thebibliography}
\end{document}